\documentclass[prl,nofootinbib,twocolumn,superscriptaddress,showpacs]{revtex4}

\usepackage{amsmath}
\usepackage{amsfonts}

\begin{document}
\begin{flushright}
http://dx.doi.org/10.1103/PhysRevLett.103.171302
\end{flushright}

\title{Constraining the energy-momentum dispersion relation\\ with Planck-scale
    sensitivity using cold atoms}

\author{Giovanni AMELINO-CAMELIA}
\affiliation{Dipartimento di Fisica, Universit\`a di Roma ``La Sapienza"\\
and Sez.~Roma1 INFN, P.le A. Moro 2, 00185 Roma, Italy}

\author{Claus L\"AMMERZAHL}
\affiliation{ZARM, Universit\"at Bremen, Am Fallturm, 28359 Bremen, Germany}

\author{Flavio MERCATI}
\affiliation{Dipartimento di Fisica, Universit\`a di Roma ``La Sapienza"\\
and Sez.~Roma1 INFN, P.le A. Moro 2, 00185 Roma, Italy}

\author{Guglielmo M. TINO}
\affiliation{Dipartimento di Fisica and LENS, Universit\`a di Firenze,\\
Sez.~INFN di Firenze, Via Sansone 1, 50019 Sesto Fiorentino, Italy}



\begin{abstract}
We use the results of ultra-precise cold-atom-recoil experiments to
constrain the form of the energy-momentum dispersion relation,
a structure that is expected to be modified in several quantum-gravity approaches.
Our strategy of analysis applies to the nonrelativistic (small speeds) limit of
the dispersion relation, and is therefore complementary to an analogous ongoing effort of
investigation of the dispersion relation in the ultrarelativistic regime using
observations in astrophysics. For the leading correction in the nonrelativistic limit the
exceptional sensitivity of cold-atom-recoil experiments remarkably allows us to set a limit
within a single order of magnitude of the
desired Planck-scale level, thereby providing the first example of Planck-scale sensitivity
in the study of the dispersion relation in controlled laboratory experiments. For the next-to-leading
term we obtain a limit which is a few orders of magnitude away from the Planck scale, but still amounts
to the best limit on a class of Lorentz-symmetry test theories that has been extensively used
to investigate the hypothesis of ``deformation" (rather than breakdown) of spacetime symmetries.
\end{abstract}

\maketitle

Gaining experimental insight on the quantum-gravity realm is very challenging,
because most effects are expected to occur on the
ultra-high ``Planck scale" $M_P (\simeq 1.2 \cdot 10^{28} eV$),
and leave only minuscule traces on processes we can access experimentally.
But thanks to a large and determined effort made over the last
decade~\cite{grbgac,astroSchaefer,astroBiller,gacNature1999,urrutiaPRL,gacQM100,jaconature,Gaclaem,PiranNeutriNat}
we do have now at least a few  research lines
in ``quantum-gravity phenomenology"~\cite{gacLRR} where
it is established that quantum properties of gravity and/or spacetime structure could be
investigated with the desired Planck-scale sensitivity.
Previously progress in this
direction had been obstructed by the extreme mathematical complexity
of the most promising theories
of quantum gravity, resulting in a
debate on quantum gravity that was confined at the level of comparison
of mathematical and conceptual features, without the ability to control the mathematics
well enough to obtain robust derivations of the physical implications of the different scenarios.

At least for one aspect of the quantum-gravity problem, the one that concerns the
possibility that spacetime itself might have to be quantized,
the nature of the debate started to change in the second half
of 1990s when it was  established that a general implication of spacetime quantization
is a modification of the classical-spacetime  ``dispersion" relation between energy $E$ and
(modulus of) momentum $p$ of a microscopic particle with mass $m$.
In the nonrelativistic limit ($p \ll m$), which is here or interest,
this dispersion relation should take the
 form
\begin{equation}
E \simeq m + \frac{p^2}{2m} + \frac{1}{2M_P}\left( \xi_1 m p + \xi_2 p^2 + \xi_3 \frac{p^3}{m} \right) ~,
 \label{DispRelNonRelativistica}
\end{equation}
working in units with speed-of-light constant set to $1$,
and including only terms
at leading order in the (inverse of
the) Planck scale.

The model-dependent dimensionless parameters $\xi_1$, $\xi_2$, $\xi_3$
should (when different from zero) have values roughly of order one, so that indeed the new effects
are introduced in some neighborhood of the Planck scale.
Evidence that at least some of these parameters should be non-zero has been found
most notably in Loop Quantum Gravity~\cite{urrutiaPRL,LQGDispRel,smolinbook},
and in particular the framework introduced in Refs.~\cite{urrutiaPRL,urrutiaPRD},
which was inspired by Loop Quantum Gravity,
produces a term linear in $p$
in the nonrelativistic limit (the effect here parametrized by $\xi_1$).
Other definite proposals for the parameters  $\xi_1$, $\xi_2$, $\xi_3$
have emerged~\cite{gacmajid,kowaPLBcosmo,Orfeupion,kodadsr}
from the quantum-gravity approach based on ``noncommutative geometry", and the associated
research area
 that contemplates deformations
of special relativity such that one could have
 an observer-indepedent maximum value of
frequency or minimum value of wavelength.
The two most studied deformation scenarios
 are the one first introduced in Ref.~\cite{gacdsrPLB2001},
whose leading-order form is
\begin{eqnarray}
E = \sqrt{m^2 +p^2} - \frac{\eta}{M_P} \frac{p^2}{2}
\label{dsr1}\end{eqnarray}
and the one first introduced in Ref.~\cite{leedsrPRL},
whose leading-order form is
\begin{eqnarray}
E = \sqrt{m^2+p^2} + \frac{\eta}{M_P} \left(\frac{m^3}{ \sqrt{m^2+p^2}}
- m^2 \right)~.
\label{dsr2}\end{eqnarray}
Interestingly both of these scenarios
have the same behaviour in the nonrelativistic
limit, dominated by a $p^2/M_P$ term of
the type here parametrized with $\xi_2$.

In addition to these examples where something definite is expected for the
parameters here of interest, which characterize the dispersion relation in the nonrelativistic
limit, there is also a quantum-gravity literature providing motivation for
studies of the dispersion relation from a broader perspective, but often within
formalisms that are not understood well enough to
establish the functional dependence of the correction on
momentum.
Nonetheless,
many authors (see, {\it e.g.}, Ref.~\cite{smolinbook} and references therein)
have argued that our best chance of having a first level of experimental characterization
of the quantum-gravity realm is through attempts to gain
insight on the parameters of the dispersion relation.

Unfortunately, as usual in quantum-gravity research, the theoretically-favoured  range
of values of the parameters of the dispersion relation translates into a range of
possible magnitudes of the effects that is extremely challenging.
If the Planck scale is the characteristic scale of quantum-gravity
effects then one expects that parameters such as $\xi_1$, $\xi_2$, $\xi_3$
should indeed take (positive or negative) values that are close to $1$,
and then, as a result of the overall factor $1/M_P$,
the effects are terribly small~\cite{gacLRR}.
Some recent semi-heuristic renormalization-group
arguments (see, {\it e.g.}, Refs.~\cite{gacLRR,wilcGUTEP} and
references therein),
have encouraged the intuition that the quantum-gravity scale might be
somewhat smaller than the Planck scale, plausibly even 3 orders of
magnitude smaller (so that it could coincide~\cite{wilcGUTEP} with
the ``grandunification scale"
which appears to be relevant in particle physics).
This would correspond to an estimate of parameters such
as $\xi_1$, $\xi_2$, $\xi_3$ plausibly as ``high" as $10^3$,
but usually even with this possible gain of 3 orders of magnitude any hope of
 detectability remains extremely distant.

It was therefore
 rather exciting for many quantum-gravity researchers
when it started to emerge that certain observations in astrophysics could
provide ``Planck-scale sensitivity" for some quantum-gravity
scenarios~\cite{grbgac,astroSchaefer,astroBiller,gacQM100,jaconature,fermiSCIENCE}.
However, these studies
only establish meaningful bounds
on scenarios with relatively strong
ultrarelativistic corrections, such as
the proposal of Ref.~\cite{gacdsrPLB2001} (Eq.~(\ref{dsr1}))
which produces a term of order $p^2/M_P$
in the ultrarelativistic regime.
But for example in the ultrarelativistic limit
of the models of Ref.~\cite{leedsrPRL} (Eq.~(\ref{dsr2}))
and Ref.~\cite{urrutiaPRL}
the effects are too small to matter.

Our main objective here is to show that cold-atom experiments
can be used to establish meaningful bounds on the parameters $\xi_1$
and $\xi_2$ that characterize the nonrelativistic limit of the dispersion
relation.
The ultra-high levels of accuracy~\cite{ChuNature02,udem} achievable with
atom interferometry have been already exploited extensively in many areas of
physics, including precision measurements of gravity \cite{Peters99}, gravity
gradients \cite{McGuirk02}, and rotation of the Earth \cite{Gustavson97Canuel06},
and also tests of Einstein's weak equivalence principle~\cite{Peters99,Fray04Dimopoulos07},
tests of Newton's law at short distances~\cite{Ferrari06},
and measurements of fundamental physical constants~\cite{mullechu, Lamporesi08}.
Clearly for our purposes it is very significant that these remarkable accuracy levels have been reached
in studies of nonrelativistic atoms.

The measurement strategy we here propose is applicable to measurements
of the ``recoil frequency" of atoms with experimental setups
involving one or more ``two-photon Raman transitions"~\cite{Kasevich91b,Peters99,Wicht02}.
Let us initially set aside the possibility of Planck-scale effects,
and discuss
the recoil of an atom in a two-photon Raman transition
from the perspective adopted
in Ref.~\cite{Wicht02},
which provides a convenient starting point for the Planck-scale generalization
we shall discuss later.
One can impart momentum to an atom through a process involving absorption
of a photon of frequency $\nu$ and (stimulated~\cite{Kasevich91b,Peters99,Wicht02})
emission, in the opposite direction,
of a photon of frequency $\nu'$. The frequency $\nu$ is computed taking into
account a resonance frequency $\nu_*$
of the atom and the momentum the atom acquires,  recoiling upon absorption
of the photon: $ \nu \simeq  \nu_* + ( h \nu_* + p)^2/(2 m) - p^2/(2m)$,
where $m$ is the mass of the atom ({\it e.g.} $m_{Cs} \simeq 124~GeV$ for Caesium),
and $p$ its initial momentum.
The emission of the photon of frequency $\nu'$ must be such to deexcite the atom
and impart to it additional
momentum: $\nu' + (2 h \nu_* + p)^2/(2 m) \simeq  \nu_* + (h \nu_*+p)^2/(2 m)$.
Through this analysis one establishes that by measuring $\Delta \nu \equiv \nu - \nu'$,
in cases (not uncommon) where $\nu_*$ and $p$ can be accurately determined, one
actually measures $h/m$ for the atoms:
\begin{eqnarray}
     \frac{\Delta \nu}{ 2 \nu_* (\nu_* +p/h)} = \frac{h}{m} ~. \label{deltaomeNOEP}
\end{eqnarray}
This result has been confirmed experimentally with remarkable accuracy.
A powerful way to illustrate this success is provided by comparing
the results of atom-recoil
measurements of $\Delta \nu/[\nu_* (\nu_* +p/h)]$ and of measurements~\cite{gab08}
of $\alpha^2$, the square
of the fine structure constant. $\alpha^2$ can be expressed in terms of the mass $m$
of any given particle~\cite{Wicht02} through the Rydberg constant, $R_\infty$,
 and the mass of the electron, $m_{{e}}$, in the following
 way~\cite{Wicht02}: $ \alpha^2 = 2 R_\infty \frac{m}{m_{{e}}} \frac{h}{m}$.
Therefore according to Eq.~(\ref{deltaomeNOEP}) one should have
\begin{equation}
\frac{\Delta \nu}{ 2 \nu_* (\nu_* +p/h)} =
\frac{\alpha^2}{2 R_\infty}
\frac{m_e}{m_u} \frac{ m_u}{m} ~, \label{alphaJ2}
\end{equation}
where $m_u$ is the
atomic mass unit and $m$ is the mass of the atoms used in
measuring $\Delta \nu/[\nu_* (\nu_* +p/h)]$. The outcomes of atom-recoil measurements,
such as the ones with Caesium reported in Ref.~\cite{Wicht02},
are consistent with Eq.~(\ref{alphaJ2})
with the accuracy
of a few parts in $10^9$.

The fact that Eq.~(\ref{deltaomeNOEP}) has been verified to such a high degree
of accuracy proves to be very
valuable for our purposes as we find
that modifications of the dispersion relation require a modification
of  Eq.~(\ref{deltaomeNOEP}). Our derivation can be summarized briefly
by observing that
the logical steps described above for the derivation of Eq.~(\ref{deltaomeNOEP})
establish the following relationship
\begin{equation}
 \Delta \nu \simeq  E(p + h\nu + h\nu') - E(p) \simeq E(2  h\nu_* + p) - E(p) ~, \label{DeltaOmegaGenerico}
\end{equation}
and therefore  Planck-scale modifications of the
dispersion relation,
parametrized in Eq.~(\ref{DispRelNonRelativistica}),
would affect $\Delta \nu$ through the modification
of $E(2 h \nu_* + p) - E(p)$, which compares the energy of
the atom when it carries momentum $p$ and when it carries momentum $p+2 h \nu_*$.

Since our main objective here is to expose sensitivity to a meaningful range of
values of the parameter $\xi_1$,
let us focus on the Planck-scale corrections
 with coefficient $\xi_1$.
In this case the relation (\ref{deltaomeNOEP}) is replaced by
\begin{eqnarray}
\Delta \nu \!  & \simeq & \! \frac{ 2 \nu_* (h \nu_* +p)}{m} +  \xi_1 \frac{m}{M_P} \nu_*
~,
\label{DeltaOmegaLeading}
\end{eqnarray}
and in turn in place of Eq.~(\ref{alphaJ2})
one has
\begin{equation}
\frac{\Delta \nu}{ 2 \nu_* (\nu_* \! + \! p/h)} \!\! \left[ \! 1
\! - \xi_1 \! \left( \! \frac{ m}{2 M_P} \! \right) \!\! \left( \! \frac{m}{h \nu_* +p} \! \right)
 \! \right] \!\! = \!\!
\frac{\alpha^2}{2 R_\infty}
\frac{m_e}{m_u} \frac{ m_u}{m} ~~~~(7b)
\nonumber
\end{equation}
We have arranged the left-hand side of this equation placing emphasis on the
fact that our quantum-gravity
correction is as usual penalized
 by the inevitable Planck-scale suppression (the ultrasmall factor $m /M_P$),
 but in this specific context it also receives a sizeable boost by the large
 hierarchy of energy scales $m/(h \nu_* +p)$, which
in typical experiments of the type here of interest can be~\cite{Kasevich91b,Peters99,Wicht02}
of order $\sim 10^{9}$.

This ``amplification" of $\sim 10^{9}$ turns out to be sufficient for our purposes:
one easily finds that, in light of our result $(7b)$,
the mentioned Caesium-atom recoil mesaurements\footnote{The Rubidium-atom recoil measurements
reported in Ref.~\cite{biraben08}
determine ${\Delta \nu}/[ 2 \nu_* (\nu_* \! + \! p/h)]$
with accuracy comparable
to the Caesium experiments of Ref.~\cite{Wicht02}. However,
in the setup of Ref.~\cite{biraben08} the Rubidium atoms had momentum $p$
significantly higher than for the Caesium atoms in Ref.~\cite{Wicht02},
and, as a consequence of the specific dependence on $p$ of our result,
it turns out that the Caesium mesaurements lead to a significantly
more stringent limit on $\xi_1$ than the Rubidium measurements.}
reported in Ref.~\cite{Wicht02},
also exploiting the high precision of a  determination
of $\alpha^2$ recently obtained from electron-anomaly measurements~\cite{gab08},
allow us to determine that $\xi_1 = - 1.8 \pm 2.1$.

From this we derive the main result we are here reporting which is
the bound $-6.0  < \xi_1 < 2.4 $,
 established at the 95\% confidence level.
This shows that the cold-atom experiments we here considered can be described as the first
example of controlled laboratory experiments probing the form of the dispersion
relation (at least in one of the directions of interest) with sensitivity that
is meaningful from a Planck-scale perspective.
We are actually already excluding a very substantial portion of the range
of values of $\xi_1$ that could be natural from a quantum-gravity perspective,
which, for reasons we briefly revisited above, goes from $|\xi_1| \sim 1$
to  $|\xi_1| \sim 10^3$.

Of course, studies of possible modifications of
the dispersion relation are also of interest for the community
involved in tests of Lorentz symmetry from a broader fundamental-physics perspective.
And our bound on the parameter $\xi_1$ is also relevant for a class
of modifications of the dispersion relation that has been studied from this
broader perspective, by introducing a parameter $\lambda$ such
that $E^2 = m^2+p^2 + 2 \lambda p$.
For this framework the previous reference limit was established in Ref.~\cite{cortes},
which considered various strategies for obtaining bounds at the level $\lambda <10~eV$.
Taking into account that from $E^2 = m^2+p^2 + 2 \lambda p$ it follows that in the
nonrelativistic limit $E = m + p^2/(2m) + \lambda p /m$, one easily finds that our
parametrization and the parametrization of Ref.~\cite{cortes} are
related by $\xi_1 m/M_P \equiv 2 \lambda/m$. And our bound on $\xi_1$ amounts
to the bound $-3.7 \cdot 10^{-6}~eV < \lambda < 1.5 \cdot 10^{-6}eV$.
From this perspective one should therefore observe that
the remarkable accuracy of cold-atom experiments allowed
us to improve on the previous best limit on $\lambda$ by more than 6 orders of magnitude!

While our main results concern indeed the parameter $\xi_1$,
we find appropriate to also briefly discuss the implications of cold-atom
studies for the term with coefficient $\xi_2$.
As mentioned, the term with coefficient $\xi_2$
in the nonrelativistic limit is a common feature of the two quantum-gravity-inpired
proposals here characterized in Eqs.~(\ref{dsr1}) and (\ref{dsr2}).
Let us notice that the same behaviour in the nonrelativistic limit
is also found in the model of Ref.~\cite{colgla},
whose proposal was not motivated by quantum gravity but has been much studied
from the broader Lorentz-symmetry-test perspective. Interestingly, for these 3 models
with the same  dependence on momentum
of the correction to energy in the nonrelativistic limit
one finds completely different consequences in the ultrarelativistic regime.
For the model of Eq.~(\ref{dsr1})
the leading ultrarelativistic correction to energy has behaviour $p^2/M_P$
and can be tightly constrained in astrophysics~\cite{grbgac,astroSchaefer,astroBiller}.
And for the model of Ref.~\cite{colgla}, whose leading ultrarelativistic correction
to energy is instead linear in momentum, a similar strategy allows to
set stringent limits using astrophysics data~\cite{colgla,steckeNEW}.
But for the third of these possibilities, the one
of Ref.~\cite{leedsrPRL} (Eq.~(\ref{dsr2})), the leading correction
to energy in the ultrarelativistic limit is only of magnitude $m^3/(p M_P)$
and cannot be significantly bounded in astrophysics.
The effort of constraining the parameter $\xi_2$ in the
nonrelativistic limit is not a top priority for the scenarios
of Ref.~\cite{gacdsrPLB2001} (Eq.~(\ref{dsr1})) and Ref.~\cite{colgla},
since those scenarios can be even more tightly constrained studying
their ultrarelativistic behaviour,
but on the contrary for the scenario
of Ref.~\cite{leedsrPRL} (Eq.~(\ref{dsr2})) the only way to establish
meaningful bounds is by investigating the nonrelativistic limit.

Following the same steps of the analysis we performed above for
the correction term with coefficient $\xi_1$,
it is easy to verify that the correction term with coefficient $\xi_2$
would produce the following modification of Eq.~(\ref{alphaJ2}):
\begin{equation}
\frac{\Delta \nu}{ 2 \nu_* (\nu_* \! + \! p/h)} \!\! \left[  1
\! - \xi_2 \frac{m}{M_P}
 \right] \!\! = \!\!
\frac{\alpha^2_*}{2 R_\infty}
\frac{m_e}{m_u} \frac{ m_u}{m}  \label{alphaJ2new2}
\end{equation}
And in this case the experimental results reported in Ref.~\cite{Wicht02}
allow us to establish that  $- 3.8 \cdot 10^{9} < \xi_2 < 1.5 \cdot 10^{9}$.
This bound is
 still some 6 orders of magnitude above even the most optimistic
 quantum-gravity estimates.
But it is a bound that still carries some significance
in the broader realm of Lorentz-symmetry investigations. According to standard
quantum-spacetime arguments, bounds on parameters such as $\xi_2$ at the level
of $|\xi_2| < 10^{9}$ amount to probing spacetime structure down to length
scales of order $10^{-26}m$ ($\sim \xi_2 h/M_P$), and, while this is not enough
for quantum gravity according to the prevailing consensus,
still represents remarkably short distance scales from a broader perspective.

Moreover our limit on $\xi_2$ at the level $|\xi_2|\lesssim 10^{9}$
indeed also
amounts to the best limit
 on the scenario
for deformation of Lorentz symmetry introduced in Ref.~\cite{leedsrPRL},
since in the nonrelativistic limit the parameter $\eta$ of Eq.~(\ref{dsr2})
is related to $\xi_2$ by $\xi_2 = 4 \eta$.
Previous attempts to constrain the parameter $\eta$ of Eq.~(\ref{dsr2})
had focused on the ultrarelativistic limit of Eq.~(\ref{dsr2}),
and did not go beyond~\cite{gacBENEDETTIdandrea,napophen}
sensitivities at the level $|\eta| \lesssim 10^{24}$.

In light of the remarkable pace of improvement of cold-atom experiments over the last 20 years,
we expect that the sensitivities here established might be improved upon in the near
future. This will most likely translate into more stringent bounds, but,
particularly considering the values of $\xi_1$ being probed,
should also be viewed as a (slim but valuable) chance for
a striking discovery. We therefore feel
that our analysis should motivate a vigorous effort on the quantum-gravity side aimed
at overcoming the mentioned technical difficulties that presently
obstruct the derivation of more detailed quantitative predictions in some of the relevant
theoretical frameworks.

\section*{Acknowledgments}
G.~A.-C. is supported in part by grant RFP2-08-02 from The Foundational
Questions Institute (fqxi.org).
C. L. aknowledges support from
the German Research Foundation and the Centre for Quantum Engeneering and Space-Time Research QUEST.

\end{document}